\documentclass[aps,pre,twocolumn,amsmath,amssymb,amsfonts,floatfix,superscriptaddress,showkeys,showpacs]{revtex4-1}
\usepackage{dcolumn}  % Align table columns on decimal point
\usepackage{multirow}
\usepackage[T1]{fontenc}
\usepackage[latin1]{inputenc}
\usepackage{dsfont}
\usepackage{verbatim}
\usepackage{bm} % bold math
\usepackage{hyperref}
\usepackage{cleveref}
\usepackage{floatrow}
\usepackage{natbib}
\usepackage{graphicx}
\usepackage{epstopdf}
\usepackage{ifpdf} % This lets us get figures right both in pdf and ps versions
\usepackage{epsfig}
\usepackage{color}
\usepackage[usenames,dvipsnames]{xcolor}
\usepackage[caption=false]{subfig}
\usepackage[export]{adjustbox}
\usepackage[colorinlistoftodos]{todonotes}
\begin{document}

\title{Giant negative mobility of inertial particles caused by the periodic potential in steady laminar flows}

\author{Bao-quan \surname{Ai}}\thanks{Email: {aibq@scnu.edu.cn}}
\affiliation{Guangdong Provincial Key Laboratory of Quantum Engineering and Quantum Materials, School of Physics and Telecommunication
Engineering, South China Normal University, Guangzhou 510006, China.}

\author{Wei-jing \surname{Zhu}}%\thanks{Email: {aibq@scnu.edu.cn}}
\affiliation{Guangdong Provincial Key Laboratory of Quantum Engineering and Quantum Materials, School of Physics and Telecommunication
Engineering, South China Normal University, Guangzhou 510006, China.}

\author{Ya-feng \surname{He}} \thanks{Email: {heyf@hbu.edu.cn}}
\affiliation{College of Physics Science and Technology, Hebei University, Baoding 071002, China}

\author{Wei-rong \surname{Zhong}} \thanks{Email: {wrzhong@jnu.edu.cn}}
\affiliation{Siyuan Laboratory, Guangzhou Key Laboratory of Vacuum Coating Technologies and
New Energy Materials, Department of Physics, Jinan University, Guangzhou 510632, China.}

\begin{abstract}
   \indent Transport of an inertial particle advected by a two-dimensional steady laminar flow is numerically investigated in the presences of a constant force and a periodic potential. Within particular parameter regimes this system exhibits absolute negative mobility, which means that the particle can travel in a direction opposite to the constant force. It is found that the profile of the periodic potential plays an important role in the nonlinear response regime. Absolute negative mobility can be drastically enhanced by applying appropriate periodic potential, the parameter regime for this phenomenon becomes larger and the amplitude of negative mobility grows exceedingly large (giant negative mobility). In addition, giant positive mobility  is also observed in the presence of appropriate periodic potential.
\end{abstract}

%\pacs{}

\maketitle

%%%%%%%%%%%%%%%%%%%%%%%%%%%%%%%%%%%%%%%%%%%%%%%%%%%%%%%%%%%%%%%%%
%%%%%%%%%%%%%%%%%%%%%%%%%%%%%%%%%%%%%%%%%%%%%%%%%%%%%%%%%%%%%%%%%
\section{Introduction}
\label{sec:intro}
\indent The problem of non-equilibrium-induced transport processes has attracted much interest in theoretical as well as experimental physics. Usually, it appears obvious that a system at rest, when perturbed by an external static force, responds by moving into the direction of that force. However, in the nonequilibrium systems, there exist some counterexamples where a particle can, instead, move in the direction opposite to the direction of the applied bias. In other words, the force-velocity characteristics of the system exhibits a passage through the origin with a negative slope as its most prominent feature. This rather surprising opposite behavior is called \emph{absolute negative mobility} (ANM), has attracted the attention of researchers in the directed transport community during the last decade \cite{RMP,PR}.

\indent  The phenomenon of ANM has been observed in a variety of setups \cite{Keay,Reimann,Speer,Speer1,Hanggi,Eichhorn,Eichhorn1,Ros,Machura,Nagel,Kostur,Spiechowicz,Hennig,Mulhern,Dandogbessi,Du,Januszewski,Ghosh,Malgaretti,Marchesoni,Savel,Sarracino,Cecconi,Cecconi1, Cividini,Haljas,Slapik,Cleuren,Chen,Li,Buceta,Mangioni,Luo}.
ANM was originally perceived as being solely due to quantum mechanical effects \cite{Keay} which may not survive in the classical limit. The first classical systems exhibiting ANM consists of interacting particles, where ANM is induced by the collective effects\cite{PR,Reimann}. Shortly afterwards, in a conveniently tailored channel with inner walls, ANM was observed where a single particle was allowed to move along meandering paths \cite{Eichhorn,Eichhorn1,Ros,Speer,Speer1,Hanggi}.  Machura and coworker \cite{Machura} found that thermal equilibrium fluctuations can induce ANM of the inertial particle in a one-dimensional periodic symmetric potential, which was confirmed in the experiment involving determination of current-voltage characteristics of the microwaved-driven Josephson junction\cite{Nagel}.  This work was extended to the cases of coloured noise\cite{Kostur}, white Poissonian noise \cite{Spiechowicz}, time-delayed feedback \cite{Hennig}, non-uniform space-dependent damping \cite{Mulhern}, and potential phase modulation\cite{Dandogbessi}.  Other examples for ANM include a vibrational motor \cite{Du}, two coupled shunted Josephson junctions \cite{Januszewski}, active Janus particles in a corrugated channel \cite{Ghosh}, and entropic electrokinetics \cite{Malgaretti}. In addition, a peculiar kind of negative mobility effect can be obtained through harmonic mixing \cite{Marchesoni,Savel}. Generally, these counter-intuitive effects are due to certain trapping mechanisms in the system, such as geometric constraints, the collective effects from particles interactions, frustration in the system, and the coupling with underlying velocity fields.

\indent Recently, Sarracino and coworkers \cite{Sarracino,Cecconi} have proposed an interesting and important model where the inertial particle advected by a two-dimensional steady laminar flow can display complex behaviors. They found that the combined action of the applied force, the particle inertia, and the velocity field can induce negative differential mobility and absolute negative mobility. However, both the parameter regime for the onset of ANM and the amplitude of the negative mobility are small, which prevents the experimental implementation for observing ANM. Therefore, how to enhance ANM in this system becomes very important. In this paper, we extend this work \cite{Sarracino,Cecconi} by applying an additional periodic potential and focus on finding how the periodic potential affects the onset of ANM. It is found that the profile of the potential can strongly affects the force-velocity behaviors. For the appropriate height of the potential, the force-velocity behaviors strongly depends on the phase difference between the stream function and the potential.  When the phase difference is $0$ or $\pi$, absolute negative mobility can be drastically enhanced, the parameter regime for ANM becomes larger and the amplitude of negative mobility grows exceedingly large (giant negative mobility). However, when the phase difference is $\pi/2$ or $3\pi/2$, the mobility is always positive, moreover giant positive mobility occurs.
%%%%%%%%%%%%%%%%%%%%%%%%%%%%%%%%%%%%%%%%%%%%%%%%%%%%%%%%%%%%%%%%%
\section{Model and Framework}

\begin{figure}
\centering
\includegraphics[width=0.5\linewidth]{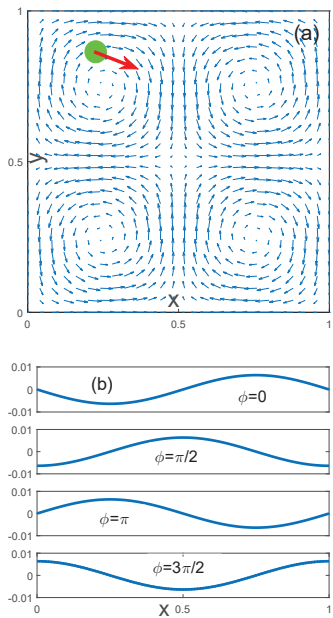}
\caption{(a) An inertial particle (green ball) moving in cellular flow field subject to a constant force $f_0$ and a periodic potential $V(x)$. Blue arrows indicate the velocity field described by Eq. (\ref{ef}). (b) The profile of the periodic potential $V(x)$ in one period described by Eq. (\ref{potential}) for different values of $\phi$. The peak of the potential is at $\frac{3}{4}L$ for $\phi=0$, $\frac{L}{2}$ for $\phi=\pi/2$, $\frac{L}{4}$ for $\phi=\pi$, and $0$ and $L$ for $\phi=3\pi/2$.}
%\label{fig:subfig}
\end{figure}

\indent We consider an inertial particle with spatial coordinates ($x$, $y$) and velocities ($v_x$, $v_y$) in a two-dimensional box of size $L\times L$ with periodic boundary conditions shown in Fig. 1(a). The particle travels through a divergenceless cellular flow $\vec{U}=(U_x,U_y)$ and is subjected to a constant force $f_0$ and a substrate force $F_p(x)$ along the $x$ direction. The dynamics of the inertial particle are described by the following equations \cite{Sarracino,Cecconi,Cecconi1}:

\begin{equation}\label{e1}
  \frac{dx}{dt}=v_x, \quad \frac{dy}{dt}=v_y,
\end{equation}

\begin{equation}\label{e2}
  \frac{dv_x}{dt}=-\frac{1}{\tau}(v_x-U_x)+\frac{1}{m}[f_0+F_p(x)]+\sqrt{2D_0}\zeta_x,
\end{equation}
\begin{equation}\label{e3}
  \frac{dv_y}{dt}=-\frac{1}{\tau}(v_y-U_y)+\sqrt{2D_0}\zeta_y,
\end{equation}
where $\tau$ is the Stokes time and $D_0$ is the noise intensity.  $\zeta_x$ and $\zeta_y$ are unit-variance Gaussian white noises with zero mean. The mass $m$ of the particle is set to 1 throughout the work.

\indent The divergenceless cellular flow $\vec{U}=(U_x,U_y)$ shown in Fig. 1(a) is defined by a stream-function $\Phi(x,y)$ as
\begin{equation}\label{ef}
  U_x=\frac{\partial  \Phi(x,y) }{\partial y}, \quad  U_y=-\frac{\partial  \Phi(x,y) }{\partial x},
\end{equation}
where $\Phi(x,y)=\frac{U_0}{k}\sin(kx)\sin(ky)$ with $k=2\pi/L$ and $U_0$ is the fluid velocity. The velocity field can be easily realized in a laboratory, e. g. ,, with two-sided lid-driven cavities \cite{Kuhlmann}, rotating cylinders \cite{Solomon} or magnetically driven vortices \cite{Tabeling}.

\indent The substrate force along $x$ direction $F_p=-\frac{\partial V(x)}{\partial x}$ arises from the following periodic potential shown in Fig. 1(b)
\begin{equation}\label{potential}
  V(x)=-\frac{V_0}{k}\sin(kx+\phi),
\end{equation}
where $V_0$ is the height of the potential. $\phi$ is the phase difference between $\Phi(x,y)$ and $V(x)$ in the $x$ direction. Note that we used the same periodicity for both $V(x)$ and $U(x, y)$ in the $x$ direction,  in this case, the potential has the most significant effect on ANM.

\indent Because the constant force $f_0$ and the substrate force $F_p(x)$ is applied to $x$ direction, we only consider transport behaviors along $x$ direction. In the asymptotic long-time regime, the average velocity $ \overline{V}_x$ and effective diffusion coefficient $D_x$ along $x$ direction are respectively defined as
\begin{equation}\label{}
  \overline{V}_x=\lim_{t\rightarrow \infty} \frac{\langle x(t)\rangle}{t}, \quad  D_x=\lim_{t\rightarrow \infty}\frac{1}{2t}[\langle x^2(t)  \rangle-\langle x(t)\rangle^2],
\end{equation}
where $\langle ...\rangle$ denotes average over trajectories of the particle with random initial conditions and noise realizations. The mobility along $x$ direction is defined as $\mu=\overline{V}_x/f_0$.

\indent As in Ref. \cite{Sarracino}, we also use times and lengths in units of $L/U_0$ and $L$, respectively. Here, we set $L=1.0$ and $U_0=1.0$, which defines a typical time scale of the flow $\tau_c=L/U_0=1$. In our simulations, the behavior of the quantities of interest can be corroborated by integration of the equations (\ref{e1}) (\ref{e2}) and (\ref{e3}) using the second-order stochastic Runge-Kutta algorithm \cite{Honeycutt}. The integration step time was chosen to be smaller than $10^{-4}$ and the total integration time was more than $10^5$. The stochastic averages were obtained as ensemble averages over $10^4$ trajectories with random initial conditions.
%%%%%%%%%%%%%%%%%%%%%%%%%%%%%%%%%%%%%%%%%%%%%%%%%%%%%%%%%%%%%%%%%
\section{Results and discussion}
\indent To investigate the effects of the periodic potential on the appearance of ANM, we mainly plot the force-velocity curves for different cases. In order to study in more details the dependence of ANM on the parameters, we present the contour plots of the average velocity as a function of the system parameters.

\begin{figure}
\centering
%\subfigure{\label{fig:subfig:a}
\includegraphics[width=0.45\linewidth]{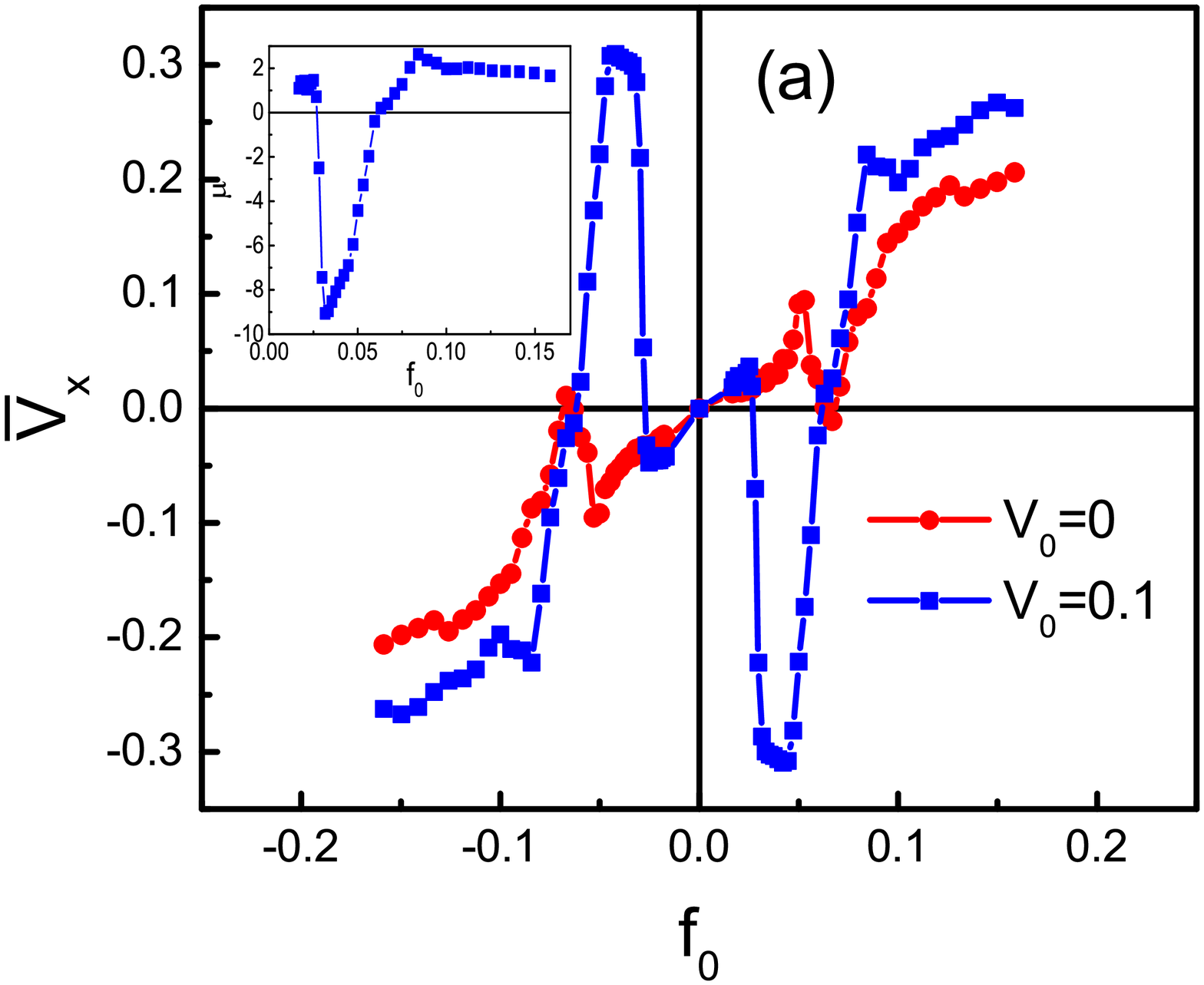}
\hspace{0.05\linewidth}
\includegraphics[width=0.45\linewidth]{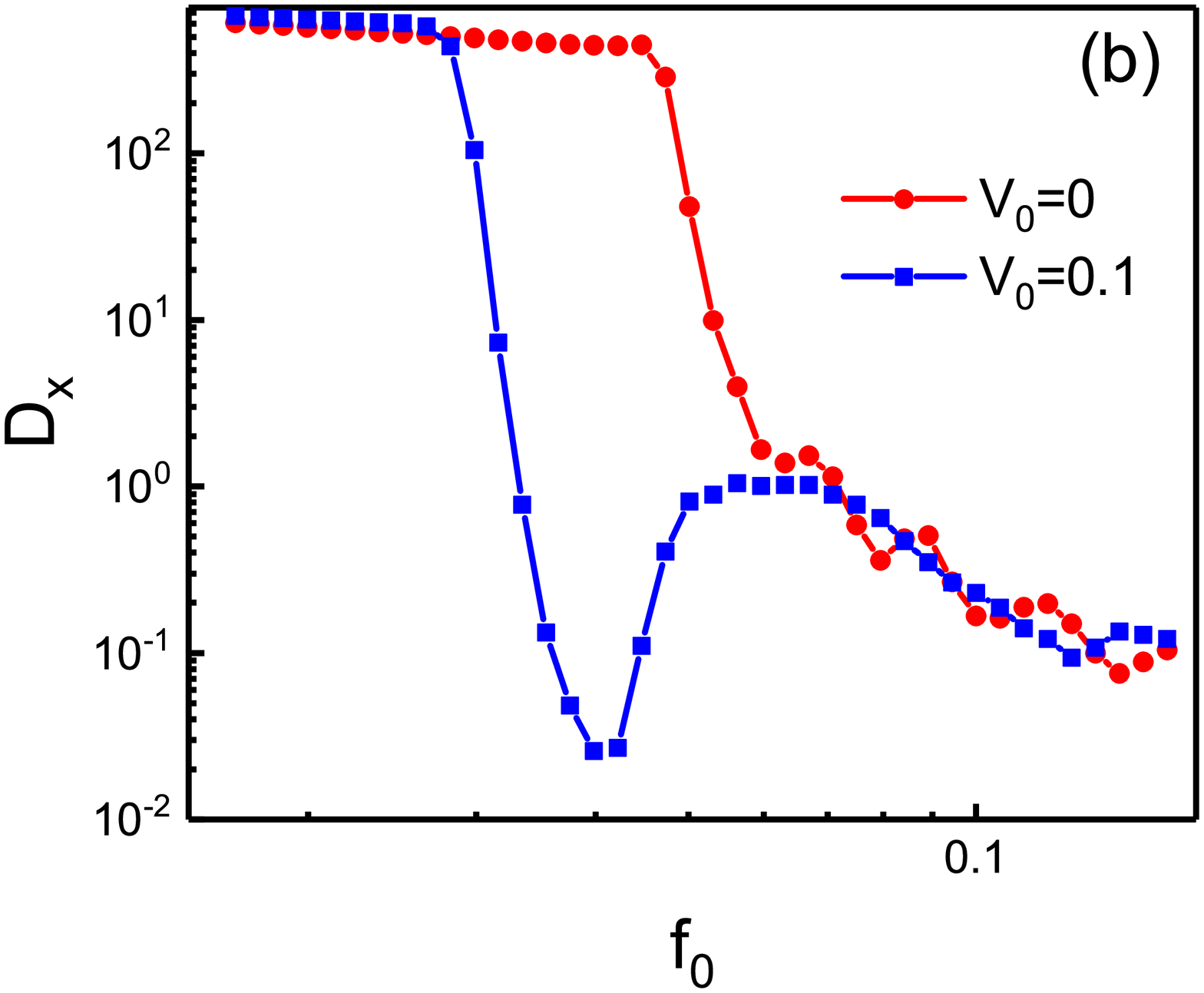}
\caption{(a) Average velocity $\overline{V}_x$ as a function of the constant force $f_0$ for $V_0=0$ and $0.1$. In the inset it is plotted the mobility $\mu$ as a function of $f_0$ for $V_0=0.1$. (b) Effective diffusion coefficient $D_x$ as a function of the constant force $f_0$ for $V_0=0$ and $0.1$. The other parameters are $D_0=10^{-5}$, $\tau=1.0$, and $\phi=0$.}
%\label{fig:subfig}
\end{figure}

\indent Figure 2(a) presents force-velocity curves with the periodic potential ($V_0=0.1$) or without the potential ($V_0=0$). The red line denotes force-velocity curve without the potential, which is the main result described in Ref. \cite{Sarracino}. In this case, a complex nonlinear behaviour can be observed for appropriate $f_0$.  There is a range of forces for which the particle has more chances to move along the channels with a direction opposite to the force. Especially, ANM (i. e. $\mu<0$) is observed in a small range around $f_0\sim \pm 0.065$ (see the red line in Fig. 2(a)). The detailed explanation for this ANM can be found in Ref. \cite{Sarracino}.  However, both the regime of ANM and the amplitude of the negative mobility are very small. When a periodic potential is applied (e. g. , $V_0=0.1$), ANM is greatly enhanced (see the blue line in Fig. 2 (a)). The regime of $f_0$ for the appearance of ANM becomes large, ANM can be observed in a range $f_0\in [0.02, 0.06]$. Moreover, for appropriate $f_0$, the amplitude of negative mobility grows exceedingly large (e. g. , $\mu=-9.0$ at $f_0=0.04$, see the curve in the inset), this phenomenon can be called as giant negative mobility \cite{Ghosh}.  The curves in Fig. 2(a) are symmetric for $f_0\rightarrow -f_0$. Therefore, we only consider the case of the positive $f_0$ in the following discussion.

\indent The effective diffusion coefficient $D_x$ as a function of $f_0$ is shown in Fig. 2(b) for different $V_0$. It is found that $D_x$ is nearly independent of $f_0$ and $V_0$ when $f_0\rightarrow 0$ or $f_0\rightarrow \infty$.  The effective diffusion coefficient for $f_0\rightarrow 0$ is much larger than that for $f_0\rightarrow \infty$ \cite{Sarracino}. For intermediate values of $f_0$, $D_x$ decreases monotonously with increase of $f_0$ without the periodic potential (i. e. $V_0=0$). However, for the case of $V_0=0.1$, on increasing $f_0$, $D_x$ firstly decreases to its minimal value, then increases to an extremum value, and finally decreases to a constant. Compared with the case of $V_0=0$, an additional peak appears at $f_0\approx 0.06$ for the case of $V_0=0.1$. This is because in the system with the tilted periodic potential the effective diffusion can be greatly enhanced near the critical bias \cite{Costantini,Reimann2}. Note that $D_x$ takes its minimal value at $f_0\approx0.04$, for which the maximal ANM is observed. Therefore, the critical bias ($f_0\approx0.06$) corresponding to the additional peak is not the same as the bias ($f_0\approx0.04$) at which the maximal ANM is observed.

\begin{figure}
\centering
%\subfigure{\label{fig:subfig:a}
\includegraphics[width=0.45\linewidth]{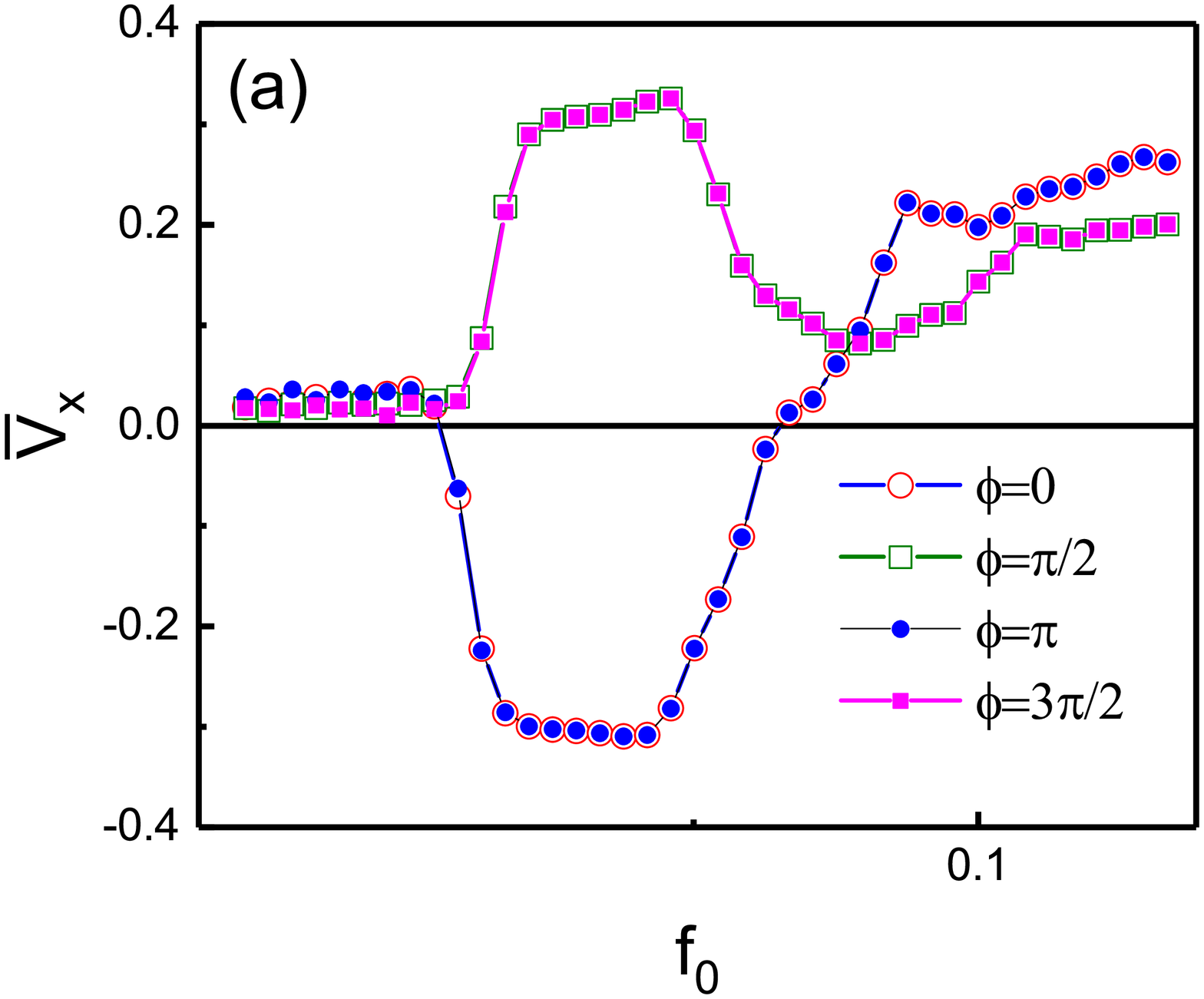}
\hspace{0.05\linewidth}
\includegraphics[width=0.45\linewidth]{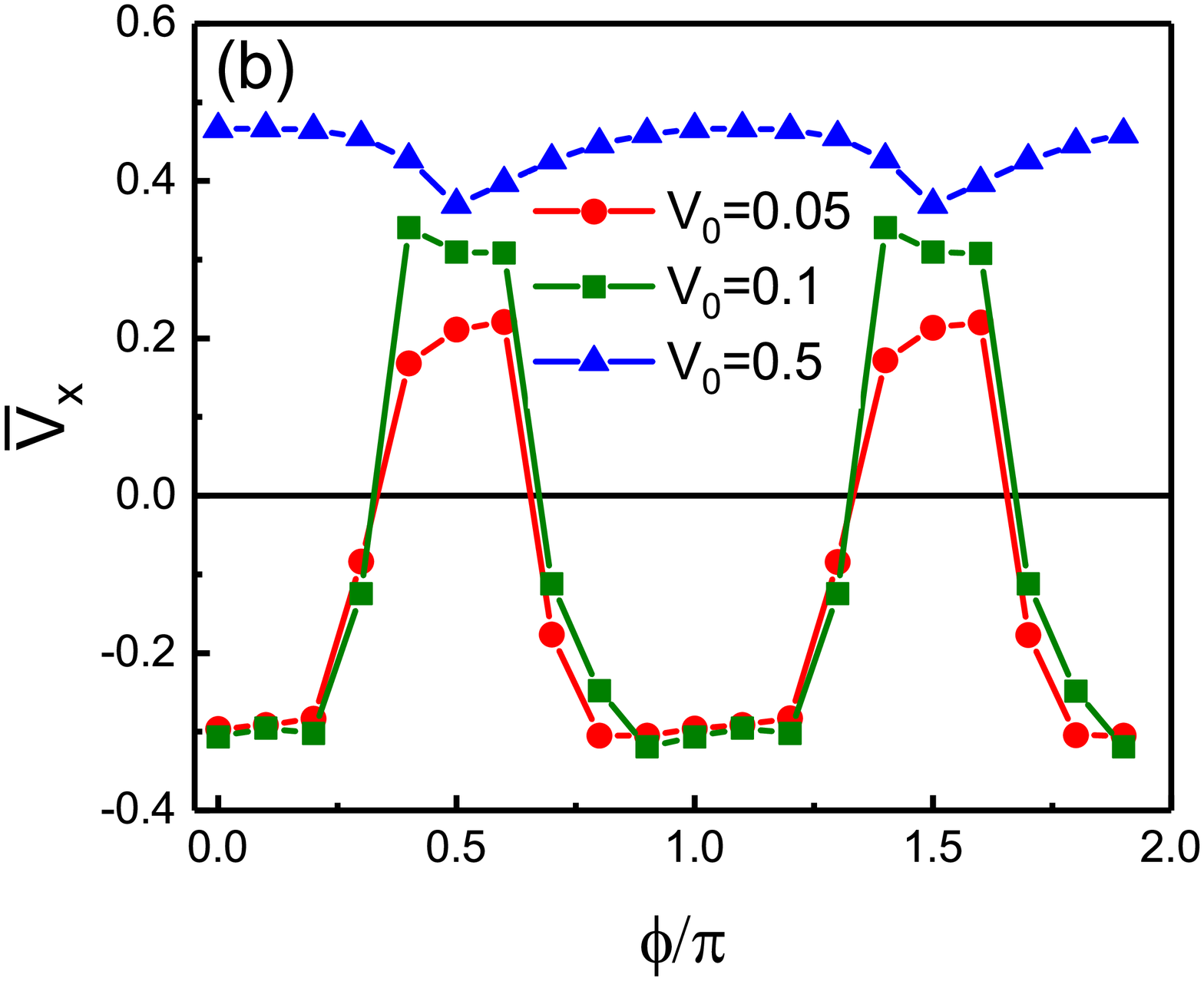}
\caption{(a) Average velocity $\overline{V}_x$ as a function of the constant force $f_0$ for different values of $\phi$ at $V_0=0.1$. (b) Average velocity $\overline{V}_x$ as a function of $\phi$ for different values of $V_0$ at $f_0=0.04$. The other parameters are $\tau=1.0$ and $D_0=10^{-5}$.}
%\label{fig:subfig}
\end{figure}
\indent Figure 3(a) shows the average velocity $\overline{V}_x$ as a function of the constant force $f_0$ for different values of $\phi$. The phase difference $\phi$ describes the profile of the potential in one period $[0, L]$. It is found that when $\phi=\pi/2$ ( or $3\pi/2$), $\overline{V}_x$ is always positive and giant positive mobility occurs. When $\phi=0$ (or $\pi$), giant negative mobility appears. The dependence of $\overline{V}_x$ on $\phi$ is plotted in Fig. 3(b).  The average velocity $\overline{V}_x$ periodically changes with the phase difference $\phi$. Especially, for appropriate $V_0$ (e. g. , $V_0=0.05$ and $0.1$), on increasing $\phi$, $\overline{V}_x$  periodically changes its direction. In one period $\phi \in [0, \pi]$, ANM will occur in the regimes $\phi\in [0, \frac{1}{4}\pi]$ and $[\frac{3}{4}\pi, \pi]$.  Therefore, the phase difference $\phi$ is very important for the appearance of ANM.

\begin{figure}
  \centering
  \includegraphics[width=0.7\linewidth]{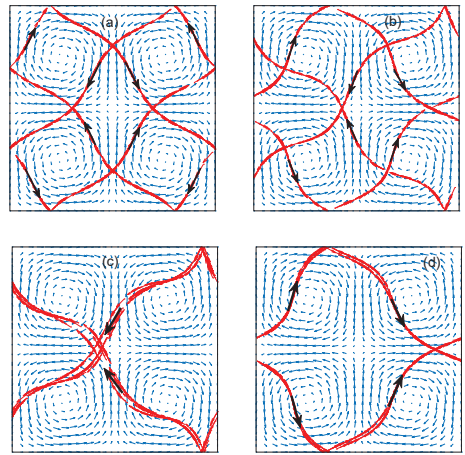}% 1\linewidth
  \caption{Sample trajectories of a particle with random initial conditions for different cases. (a) $f_0=0$ and $V_0=0$. (b) $V_0=0$ and $f_0=0.04$. (c) $V_0=0.1$, $f_0=0.04$, and $\phi=0$ or $\pi$. (d) $V_0=0.1$, $f_0=0.04$, and $\phi=\pi/2$ or $3\pi/2$. Blue arrows illustrate the underlying velocity field and red lines describes the preferential channels. The other parameters are $D_0=10^{-5}$ and $\tau=1.0$.}\label{fig.2}
\end{figure}

\indent The main result above is that ANM can drastically be enhanced by applying the periodic potential and producing a giant negative mobility.  In order to get insight into the origin of ANM, we have plotted the trajectories of inertial particles in Fig. 4.  It is found that the motion of particles is realized along some preferential channels \cite{Sarracino}. These channels fall into two categories, the rightward channel with $\overline{V}_x>0$ and the leftward channel with $\overline{V}_x<0$. Therefore, the direction of transport depends on which type of channel is dominant. Note that these preferential channels will disappear when increasing $\tau$ and $D_0$. When $V_0=0$ and $f_0=0$ (shown in Fig. 4(a)), these two types of channels are exactly the same, thus, no direction transport occurs and $\overline{V}_x=0$. When a constant force $f_0=0.04$ is applied to the particle (shown in Fig. 4(b)), the occupation of the leftward channels reduces and the rightward channels dominate the transport, particles on average move to the right, thus $\overline{V}_x>0$. Note that when $f_0=0.065$ (the ANM case in Ref. \cite{Sarracino}), there exists the probability of transitions from the rightward channels to the leftwards channels, which can induce ANM.

\indent The role of the periodic potential on ANM (shown in Figs. 4(c) and 4(d)) can be analyzed by the sample trajectories of inertial particles at $f_0=0.04$ and $V_0=0.1$ for different $\phi$.  When $\phi=0$ ($\pi$) (see the first and third panels in Fig. 1(b)), the peak (valley) of the potential is on the right and the valley (peak) is on the left, there exists a potential difference between the left and the right, which facilitates the transition between  preferential channels (specifically, from the rightward channels to the leftwards channels when $f_0=0.04$). Compared with Fig. 4(b), in Fig. 4(c) the rightward channels completely disappear and there is only the leftward channels, thus large negative velocity occurs (giant negative mobility shown in Fig. 2(a)). When $\phi=\pi/2$ ($3\pi/2$) (see the second and fourth panels in Fig. 1(b)), the peak (valley) of the potential is in the middle of a period $[0,L]$, which suppresses the transition between channels (specifically, from the rightward channels to the leftwards channels when $f_0=0.04$).  Therefore, the leftward channels completely disappear and only the rightward channels exist, thus large positive velocity occurs (giant positive mobility shown in Figs. 3(a) and 3(b)).  The profile of the potential, i. e., the phase difference $\phi$, strongly affects the onset of ANM. In the following, we only consider the case of $\phi=0$, where the phenomenon of ANM is most prominent for appropriate $V_0$.

\begin{figure*}[htbp]
  \centering
  \includegraphics[width=0.3\linewidth]{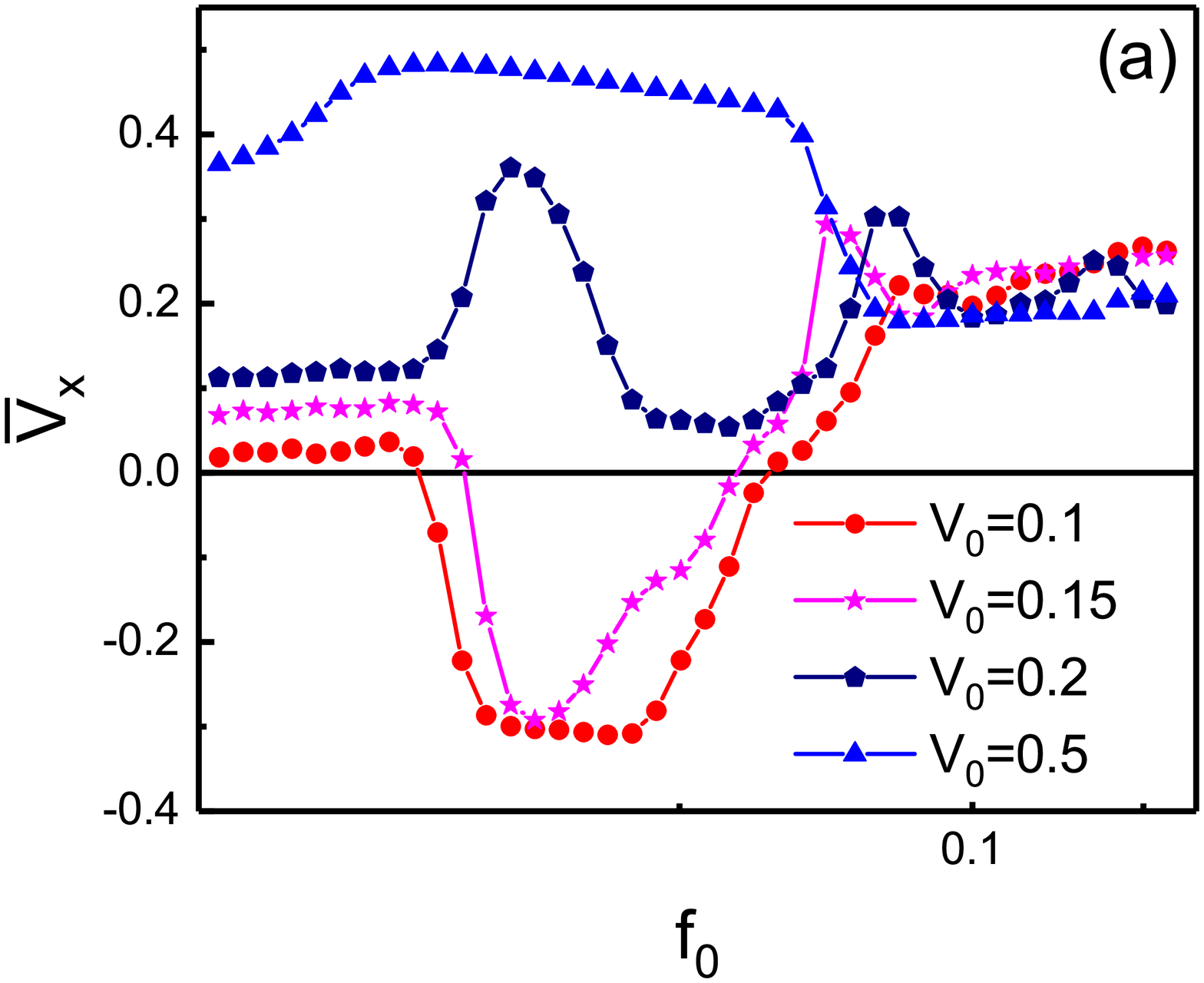}% 1\linewidth
  \hspace{0.03\linewidth}
  \includegraphics[width=0.3\linewidth]{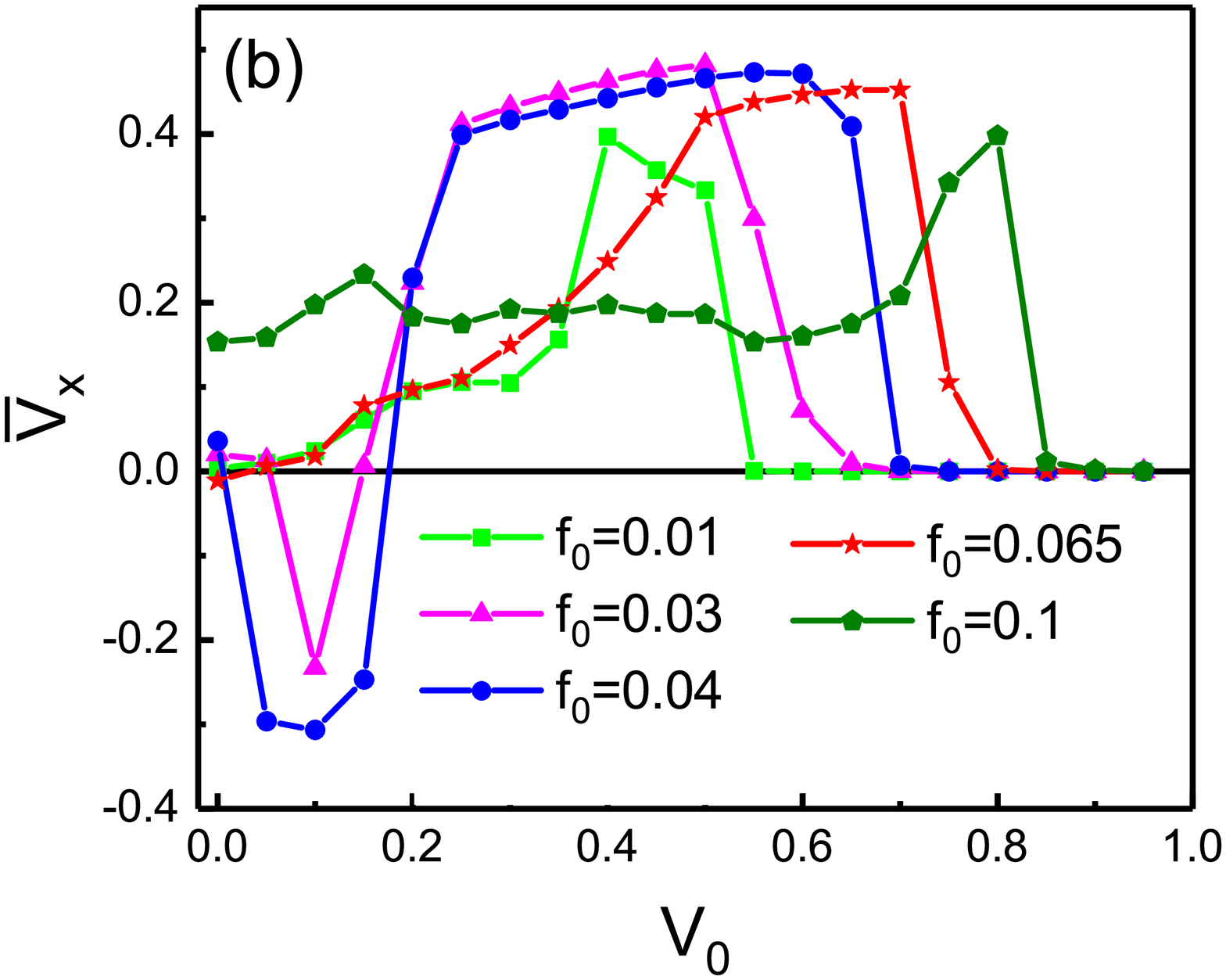}
  \hspace{0.03\linewidth}
  \includegraphics[width=0.3\linewidth]{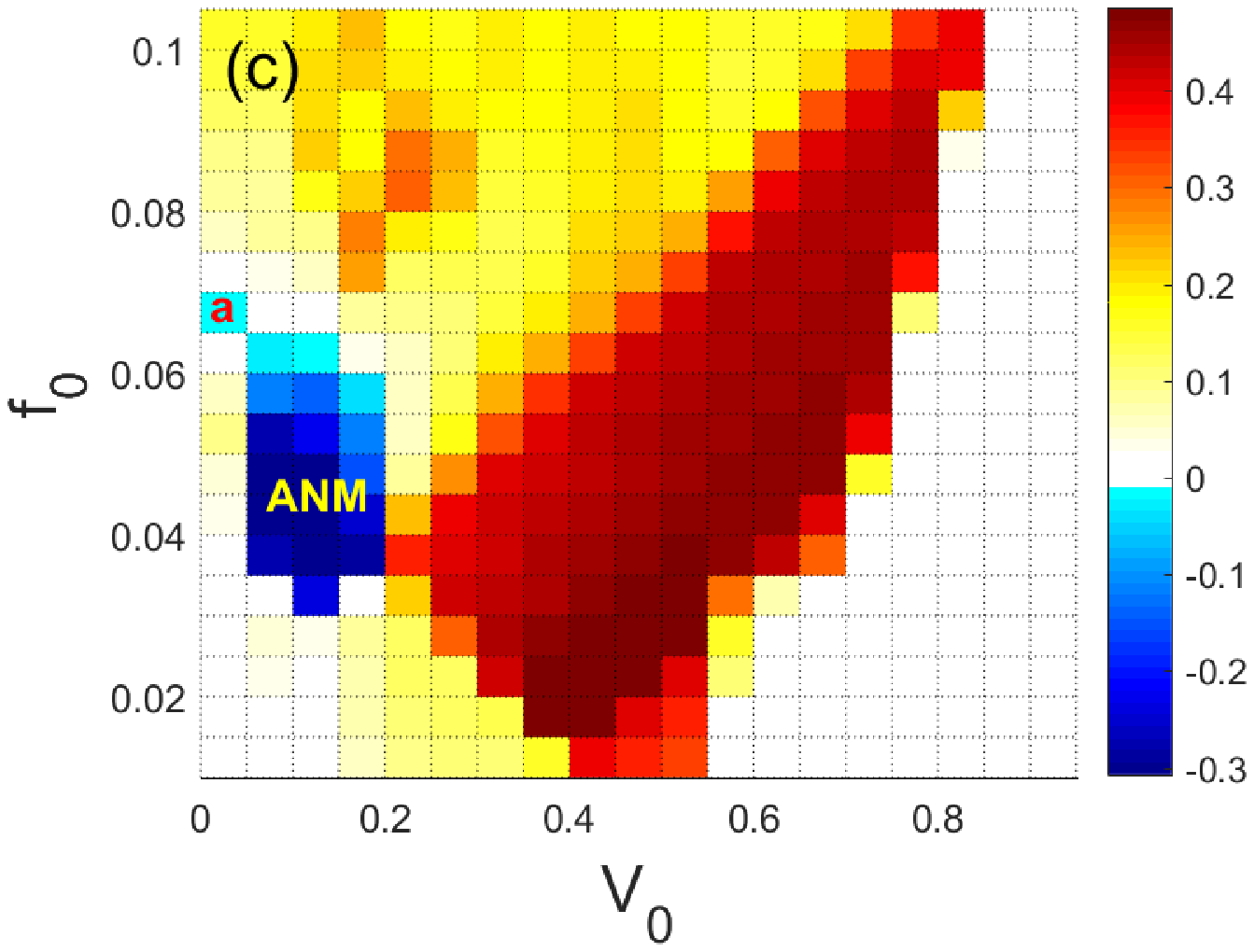}
  \caption{(a) Average velocity $\overline{V}_x$ as a function of the constant force $f_0$ for different values of $V_0$. (b) Average velocity $\overline{V}_x$ as a function of $V_0$ for different values of $f_0$. (c) Contour plots of the average velocity $\overline{V}_x$ as a function of the system parameters $V_0$ and $f_0$. The other parameters are $D_0=10^{-5}$, $\tau=1.0$, and $\phi=0$.}\label{fig.2}
\end{figure*}

\indent Figure 5(a) shows the effect of the height of the potential on the force-velocity relation when $\tau=1.0$, $\phi=0$, and $D_0=10^{-5}$. It is found that the height of the potential $V_0$ has a strong impact on the force-velocity relation.  For appropriate value of $V_0$ (e. g. ,  $V_0=0.1$ and $0.15$),  ANM can be observed in a certain range of $f_0$. When increasing $V_0$, the regime of ANM gradually becomes smaller and finally disappears for large values of $V_0$ (e. g. , $V_0=0.2$ and $0.5$). Although ANM disappears for large $V_0$, the phenomenon of negative differential mobility appears, which is characterized by a maximum for a certain value of $f_0$. Especially, when $V_0=0.2$, there exist three regimes of $f_0$, where the phenomenon of negative differential mobility appears.

\indent Figure 5(b) describes $\overline{V}_x$ versus the height $V_0$ of the potential for different $f_0$.  For very small or very large $f_0$ (e. g. , $f_0=0.01$ or $0.1$), $\overline{V}_x$ is always positive. For intermediate values of $f_0$, current reversals occur by changing $V_0$.  For the case of $f_0=0.065$ \cite{Sarracino}, on increasing $V_0$ from zero, $\overline{V}_x$ transits from negative to positive. Remarkably, for appropriate values of $f_0$ (e. g. , $f_0=0.03$ and $0.04$), $\overline{V}_x$ changes its direction twice (multiple current reversals) when $V_0$ increases from zero. Note that $\overline{V}_x$ quickly drops to zero when $V_0$ is larger than the critical value (this critical value increases with $f_0$). To study in more detail the dependence of $\overline{V}_x$ on $f_0$ and $V_0$, the contour plots of average velocity $\overline{V}_x$ as a function of the system parameters $V_0$ and $f_0$ are shown in Fig. 5(c).  The regime of ANM  marked with the letter 'a' is the results described in Ref. \cite{Sarracino} where $V_0=0$. It is found that the phenomenon of ANM will appear  in the regime of $f_0\in [0.03, 0.07]$ and $V_0\in [0, 0.2]$ and in the other regime $\overline{V}_x$  is always positive. Note that the regime of ANM will change when the other parameters are varied.

\begin{figure*}[htbp]
\centering
%\subfigure{\label{fig:subfig:a}
\includegraphics[width=0.3\linewidth]{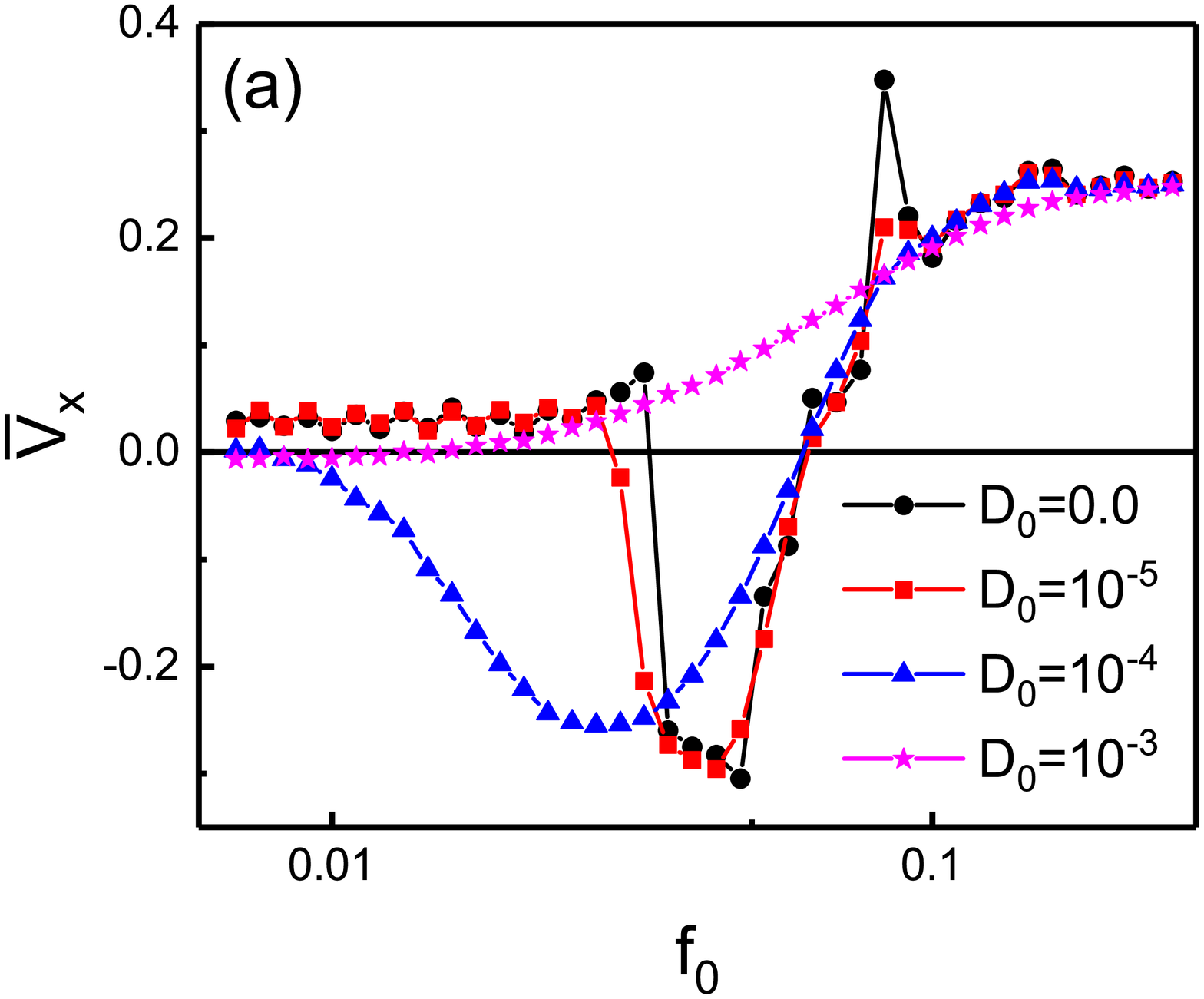}
\hspace{0.03\linewidth}
\includegraphics[width=0.3\linewidth]{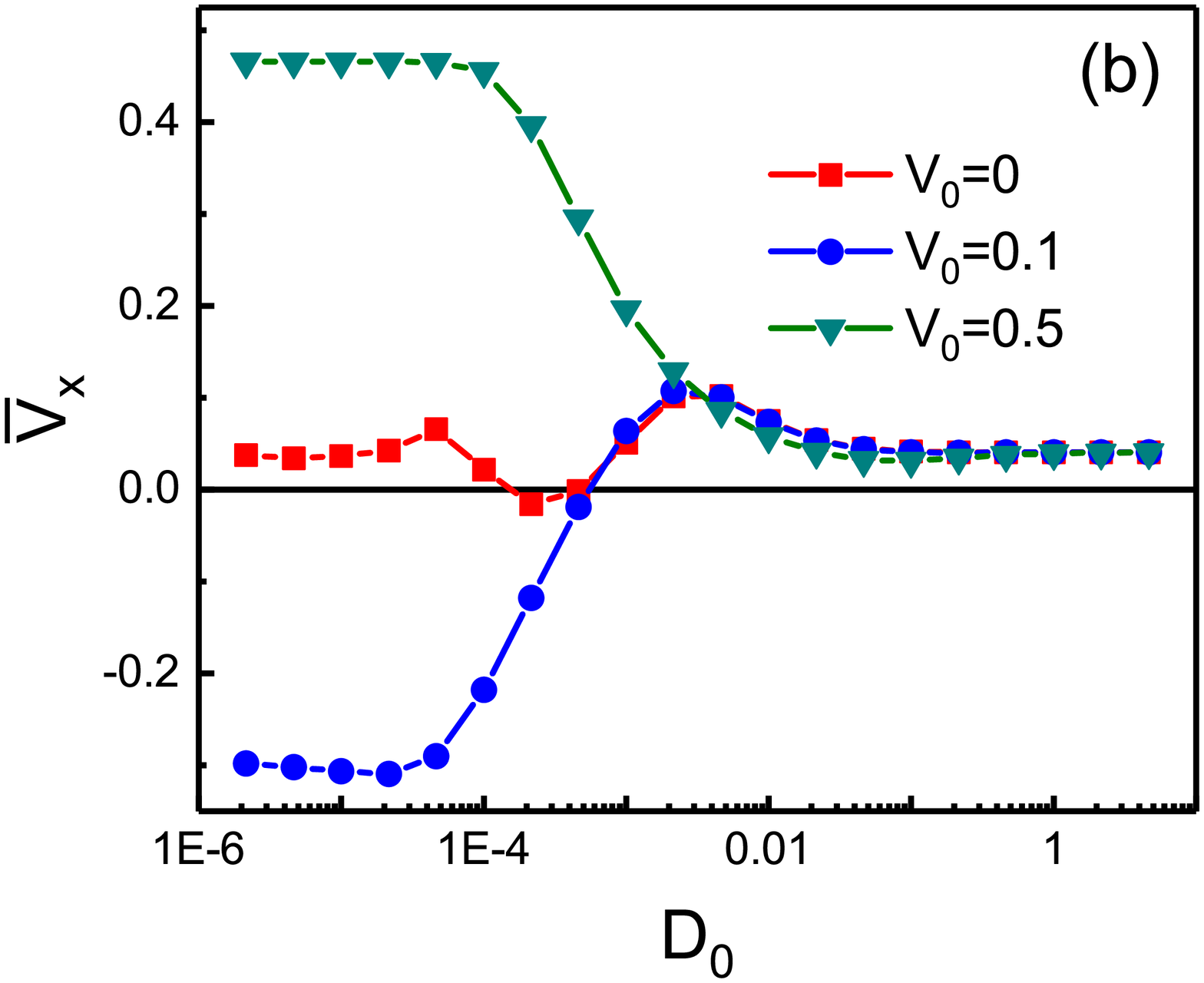}
\hspace{0.03\linewidth}
\includegraphics[width=0.3\linewidth]{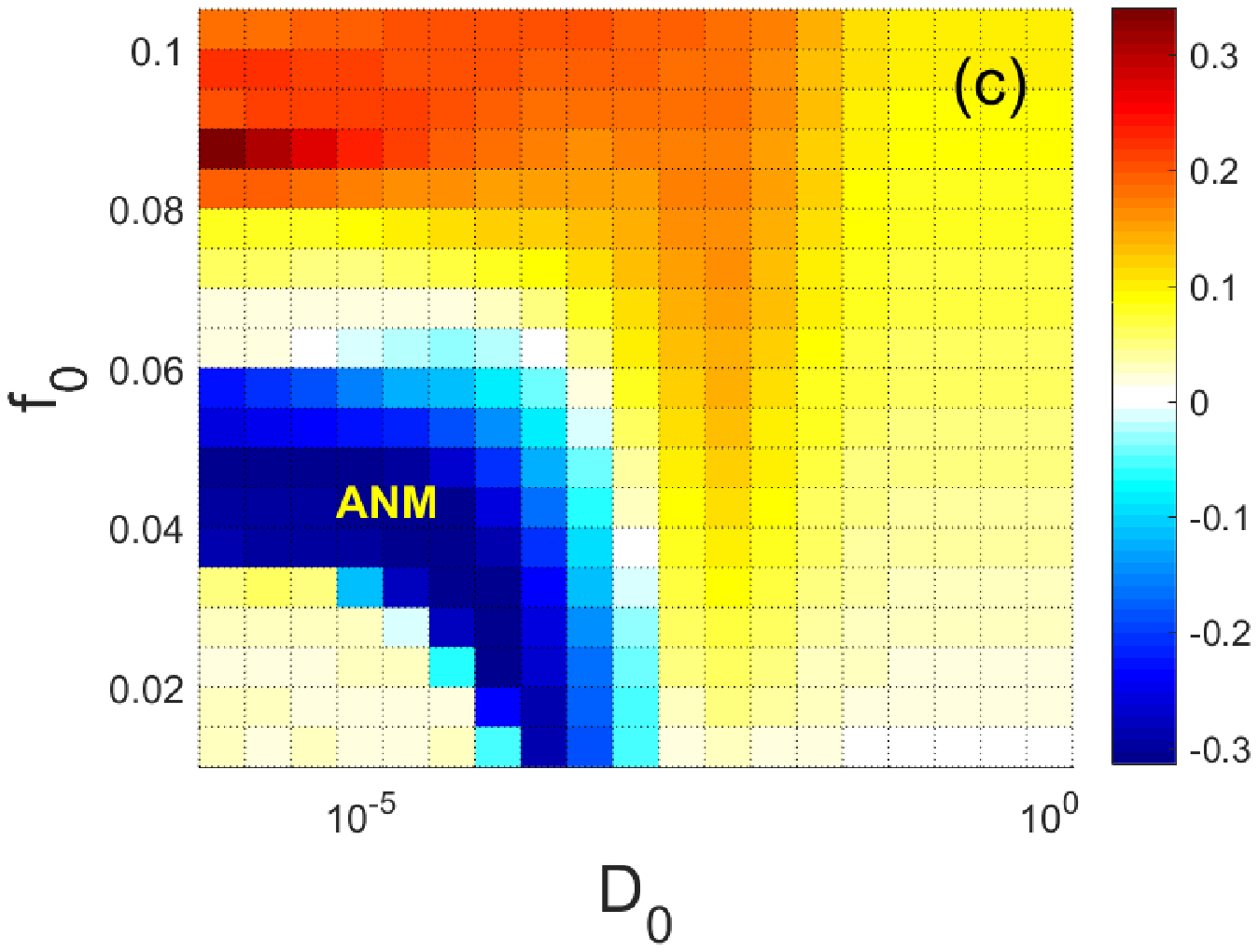}
\caption{(a) Average velocity $\overline{V}_x$ as a function of the constant force $f_0$ for different values of $D_0$. (b) Average velocity $\overline{V}_x$ as a function of $D_0$ for different values of $V_0$ at $f_0=0.04$. (c) Contour plots of the average velocity $\overline{V}_x$ as a function of the system parameters $D_0$ and $f_0$. The other parameters are $V_0=0.1$, $\tau=1.0$, and $\phi=0$.}
%\label{fig:subfig}
\end{figure*}

\indent In Fig. 6(a), we investigate the behaviors of the force-velocity curve for different values of $D_0$. We observe very obvious phenomenon of ANM when $D_0=0.0$, $10^{-5}$, and $10^{-4}$, while the phenomenon of ANM almost disappears for $D_0=10^{-3}$. Remarkably, in the case of $D_0=10^{-4}$, negative mobility appears even for small forces. This effect is due to the nonequlibrium nature of the velocity filed \cite{Sarracino,Cecconi}.  Note that the phenomenon of ANM still exists even in the absence of thermal equilibrium fluctuations (e. g. , $D_0=0$), which is different from the case in Ref. \cite{Machura}, where ANM disappears in the absence of thermal equilibrium fluctuations.

\indent The dependence of $\overline{V}_x$ on $D_0$ is plotted in Fig. 6(b) at $f_0=0.04$. When the periodic potential is absent (i. e., $V_0=0$), there exists a very narrow range of $D_0$, where ANM occurs, the amplitude of negative mobility is very small.  However, for the  appropriate $V_0$ (e. g. , $V_0=0.1$), the range of $D_0$ for ANM is greatly expanded. Moreover, giant negative mobility is observed when $D_0<10^{-4}$. When $D_0\rightarrow \infty$, the average velocity $\overline{V}_x$ tends to a constant which is independent of $V_0$.  This is expected because if the noise is strong enough the effect of the velocity field is averaged out and only the constant force dominates the transport. The regime of parameter space ($D_0$, $f_0$) for the onset of ANM is shown in Fig. 6(c) at $V_0=0.1$. On increasing $D_0$ from $10^{-6}$ to $10^{-4}$, the regime of $f_0$ for the onset of ANM changes from [$0.03$, $0.06$] to [$0.01$, $0.07$]. On further increasing $D_0$ from $10^{-4}$, the regime of $f_0$ for the onset of ANM reduces and finally disappears. Note that the average velocity $\overline{V}_x$ is always positive when $D_0>10^{-3}$ or $f_0>0.07$.

\begin{figure*}[htbp]
\centering
%\subfigure{\label{fig:subfig:a}
\includegraphics[width=0.3\linewidth]{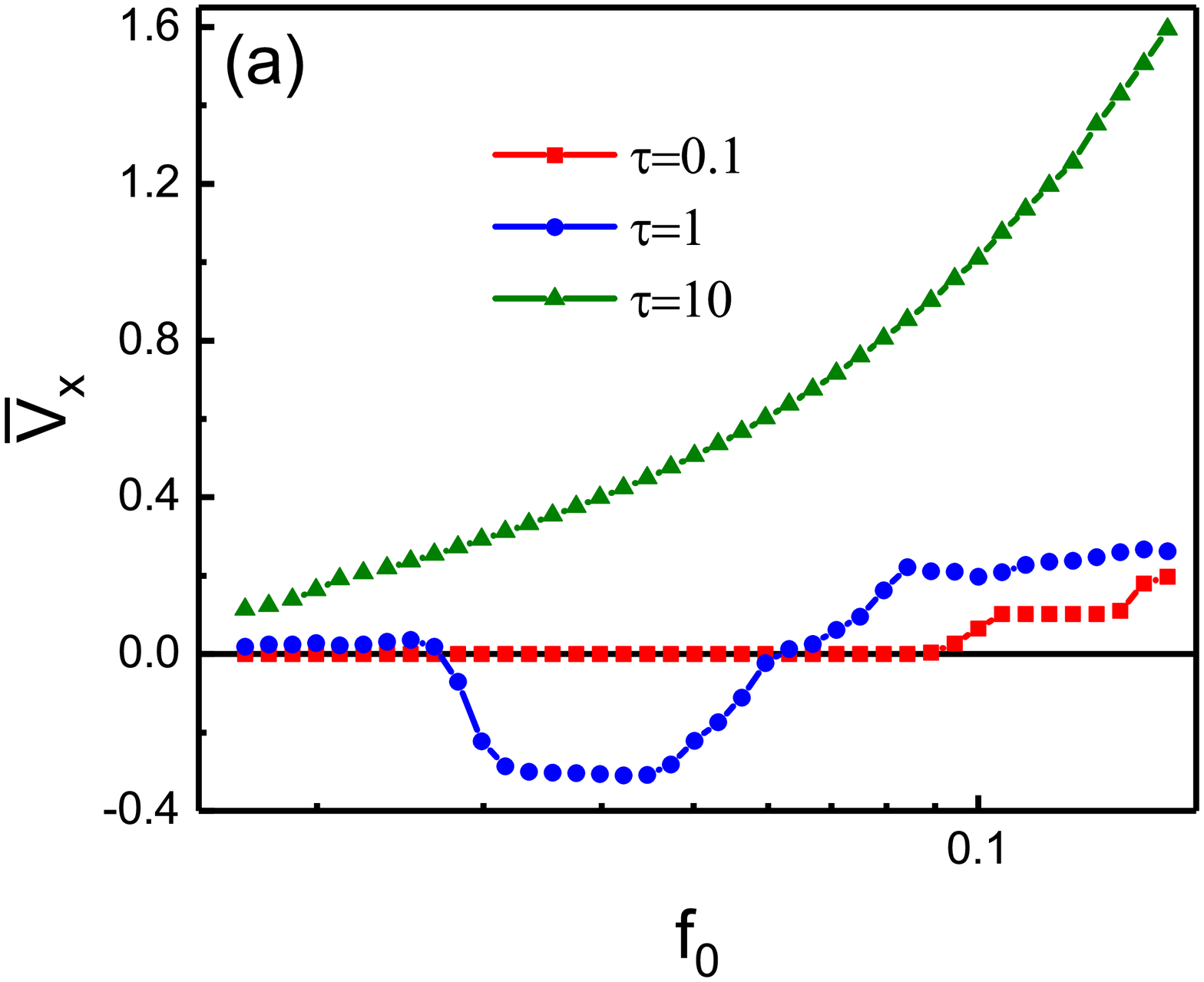}
\hspace{0.03\linewidth}
\includegraphics[width=0.3\linewidth]{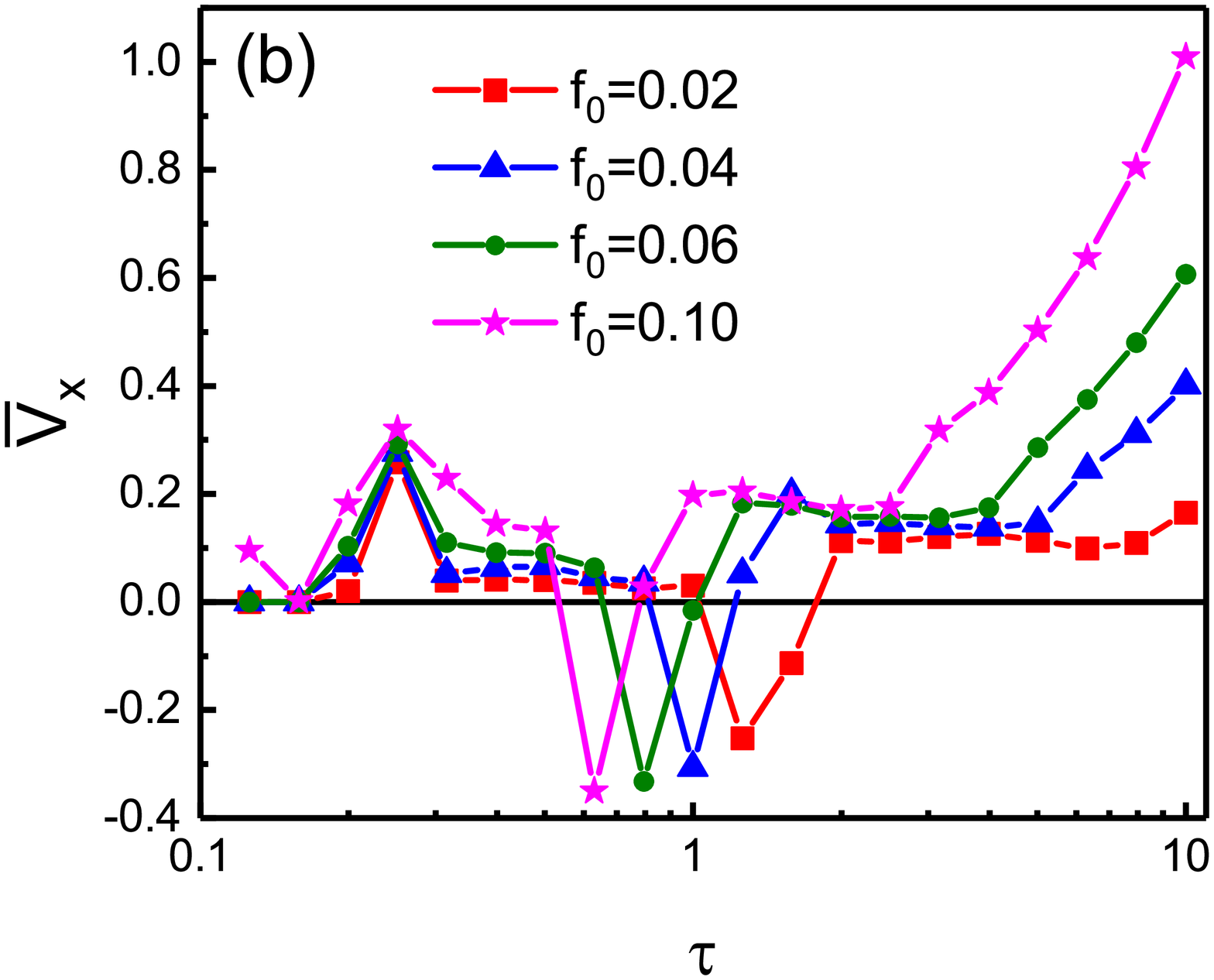}
\hspace{0.03\linewidth}
\includegraphics[width=0.3\linewidth]{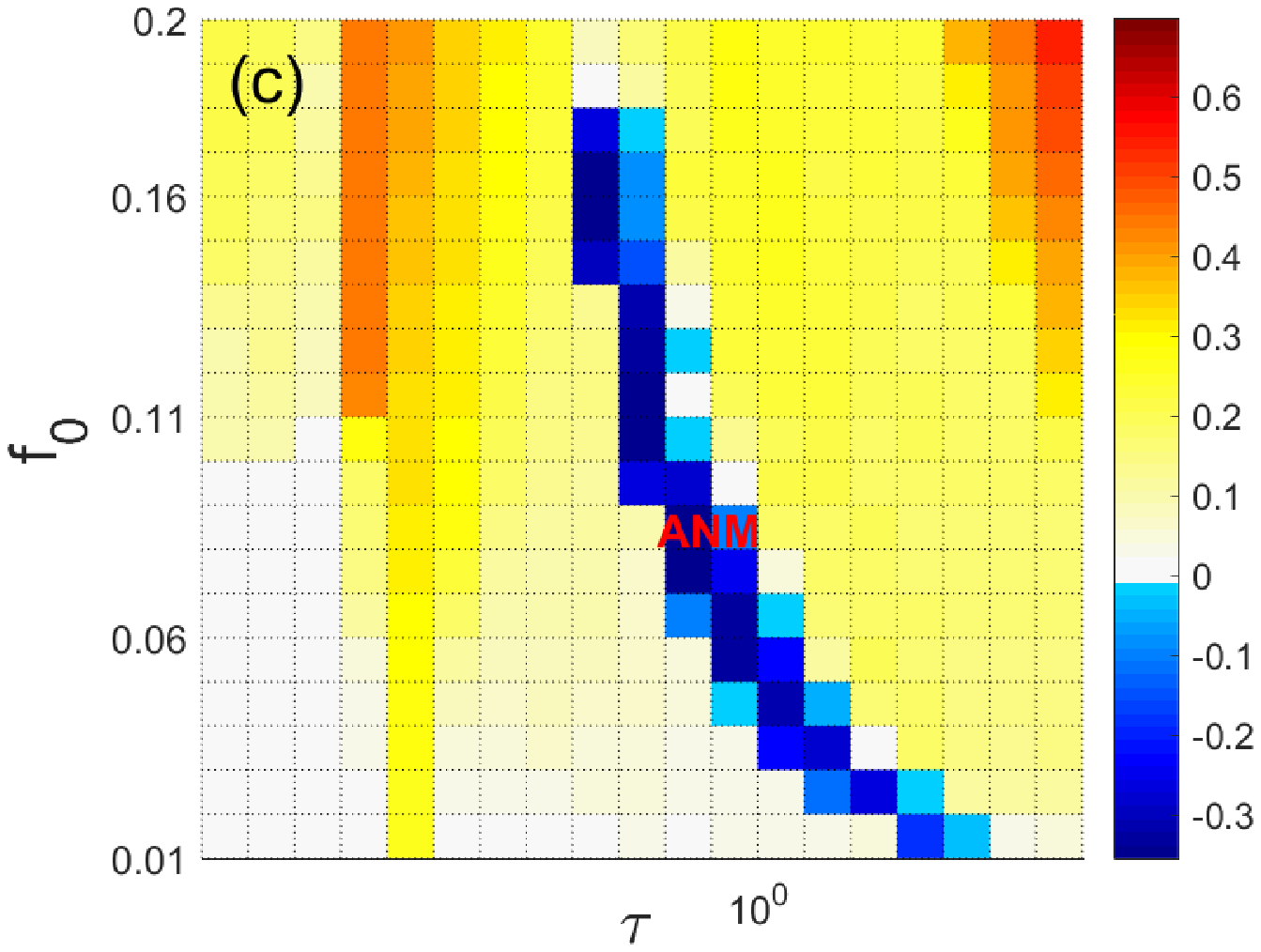}
\caption{(a) Average velocity $\overline{V}_x$ as a function of the constant force $f_0$ for different values of $\tau$. (b) Average velocity $\overline{V}_x$ as a function of $\tau$ for different values of $f_0$. (c) Contour plots of the average velocity $\overline{V}_x$ as a function of the system parameters $D_0$ and $f_0$. The other parameters are $V_0=0.1$, $D_0=10^{-5}$, and $\phi=0$.}
%\label{fig:subfig}
\end{figure*}
\indent Figure 7(a) shows the force-velocity relation for different values of $\tau$ at $V_0=0.1$, $D_0=10^{-5}$, and $\phi=0$.  For small $\tau$ (e. g. , $\tau=0.1$), the velocity filed dominates the transport, the particle is trapped in one period when $f_0<0.09$. When $f_0>0.09$, the average velocity is positive and increases with $f_0$. For large $\tau$ (e. g. , $\tau=10.0$), the velocity field is negligible, the constant force dominates the transport, thus $\overline{V}_x$ increases monotonously with $f_0$.  When $\tau$ is equal to the typical time scale $\tau_c$ ($\tau=\tau_c=1$), the phenomenon of ANM is observed in a range $f_0\in [0.02, 0.06]$.

\indent The average velocity $\overline{V}_x$ as a function of $\tau$ is plotted in Fig. 7(b) for different values of $f_0$.  For small $\tau$ (e. g. , $\tau<0.3$), $\overline{V}_x$  is positive and there exists a peak in the curve and the position of the peak is almost unchanged when $f_0$ varies.  For large $\tau$ (e. g. , $\tau>2.5$), the average velocity $\overline{V}_x$ is always positive and increases  monotonously with $\tau$. For intermediate $\tau$, the phenomenon of ANM is observed, there exists a valley in the curve and the position of the peak shifts to small $\tau$ when $f_0$ increases.  To study in more detail the dependence of the average velocity on $\tau$ and $f_0$, we plots the contour plots of $\overline{V}_x$ as a function of the system parameters $\tau$ and $f_0$ in Fig. 7(c). It is found that ANM may occur in a range $\tau\in [0.5, 2.5]$. When $\tau$ increases from $0.5$ to $2.5$, the constant force $f_0$ for the appearance of ANM decreases. Therefore, how to choose appropriate $\tau$ is very important to observe the phenomenon of ANM.

%\begin{figure}
%\centering
%\subfigure{\label{fig:subfig:a}
%\includegraphics[width=0.9\linewidth]{d0_f0.eps}
%\label{fig:subfig}
%\end{figure}

%%%%%%%%%%%%%%%%%%%%%%%%%%%%%%%%%%%%%%%%%%%%%%%%%%%%%%%%%%%%%%%%%
\section{Concluding Remarks}
\indent Based on recent work \cite{Sarracino}, we numerically studied the transport of the inertial particle driven by the two-dimensional steady laminar flow in the presences of a constant force and a periodic potential. We focus on finding how the periodic potential affects the appearance of ANM. The parameter space for the onset of ANM is very small when the periodic potential is absent. When the periodic potential is applied to the particle, the nonlinear response of the particle in the flows has changed dramatically.  The profile of the potential strongly affect the appearance of ANM. When $\phi=0$ ( or $\pi$) and $V_0=0.1$, giant negative mobility occurs (e. g. , $\mu=-9.0$ at $f_0=0.04$). Moreover, the regime of $f_0$ for the appearance of ANM becomes large, ANM can be observed in a range  $f_0\in [0.02, 0.06]$.  When $\phi=\pi/2$ (or $3\pi/2$) and $V_0=0.1$, giant positive mobility occurs.  The height of the potential $V_0$ also has a strong impact on the force-velocity relation. When increasing $V_0$, the regime of ANM gradually becomes smaller and finally disappears for large $V_0$. The nonlinear response behaviors of the inertial particle is determined by the combined action of the applied force, the particle inertia, the velocity field, and the periodic potential.

\indent  The present work do not consider the interactions between particles. It would be interesting to study how the interactions (both steric interaction and alignment interaction) and collective behaviors affect the onset of ANM in steady laminar flows. In addition, the phenomena we have presented could be experimentally observed in steady laminar flows (which can be realized in setups with rotating cylinders), where a spherical silica particle move in a periodic potential. The typical Stokes time for the onset of ANM can be  estimated $\tau=\frac{m}{6\pi\eta a}=\frac{U_0}{L}$, where $\eta$ is the solvent viscosity and $a$ is the particle radius.

%%%%%%%%%%%%%%%%%%%%%%%%%%%%%%%%%%%%%%%%%%%%%%%%%%%%%%%%%%%%%%%%%
\section*{Acknowledgements}
\indent This work was supported in part by the National Natural Science Foundation of China (Grant No. 11575064), the GDUPS (2016), the Natural Science Foundation of Guangdong Province (Grant Nos. 2017A030313029 and 2016A030313433), and the Major Basic Research Project of Guangdong Province (No. 2017KZDXM024).


\begin{thebibliography}{99}
\bibitem{RMP} P. H\"{a}nggi and F. Marchesoni, Rev. Mod. Phys. \textbf{81}, 387 (2009).
\bibitem{PR} P. Reimann, Phys. Rep. \textbf{361}, 57 (2002).
\bibitem{Keay} B. J. Keay, S. Zeuner, S. J. Allen, K. D. Maranowski, A. C. Gossard, U. Bhattacharya, and M. J. W. Rodwell, Phys. Rev. Lett. \textbf{75}, 4102 (1995).

\bibitem{Reimann} P. Reimann, C. Van den Broeck, and R. Kawai, Phys. Rev. E \textbf{60}, 6402 (1999).

\bibitem{Speer} D. Speer, R. Eichhorn, M. Evstigneev, and P. Reimann, Phys. Rev. E \textbf{85}, 061132 (2012).
\bibitem{Speer1} D. Speer, R. Eichhorn, and P. Reimann, Phys. Rev. Lett. \textbf{102}, 124101 (2009).
\bibitem{Hanggi} P. Hanggi, F. Marchesoni, S. Savel'ev, and G. Schmid, Phys. Rev. E \textbf{82}, 041121 (2010).
\bibitem{Eichhorn} R. Eichhorn,  J. Regtmeier,  D. Anselmetti,  and  Peter Reimann, Soft Matter \textbf{6}, 1858 (2010).
\bibitem{Eichhorn1} R. Eichhorn, P. Reimann, and P. H\"{a}nggi, Phys. Rev. Lett. \textbf{88}, 190601 (2002).
%\bibitem{Eichhorn2} R. Eichhorn, P. Reimann, and P. H\"{a}nggi, Phys. Rev. E \textbf{66}, 066132 (2002).
\bibitem{Ros} A. Ros, R. Eichhorn, J. Regtmeier, T. T. Duong, P. Reimann, and D. Anselmetti, Nature \textbf{436}, 928 (2005).

\bibitem{Machura} L. Machura, M. Kostur, P. Talkner, J. Luczka, and P. H\"{a}nggi, Phys. Rev. Lett. \textbf{98}, 040601 (2007).
\bibitem{Nagel} J. Nagel, D. Speer, T. Gaber, A. Sterck, R. Eichhorn, P. Reimann, K. Ilin, M. Siegel, D. Koelle, and R. Kleiner, Phys. Rev. Lett. \textbf{100}, 217001 (2008).

\bibitem{Kostur} M. Kostur, J. Luczka, P. H\"{a}nggi, Phys. Rev. E \textbf{80}, 051121 (2009).
\bibitem{Spiechowicz} J. Spiechowicz, P. H\"{a}nggi, and J. Luczka, Phys. Rev. E \textbf{90}, 032104 (2014).
%\bibitem{Spiechowicz1} J. Spiechowicz, J. Luczka, and P. Hanggi,  J. Stat. Mech.: Theor. Exp. \textbf{2013}, P02044 (2013).
\bibitem{Hennig} D. Hennig, Phys. Rev. E \textbf{79}, 041114 (2009).
\bibitem{Mulhern} C. Mulhern, Phys. Rev. E \textbf{88}, 022906 (2013).
\bibitem{Dandogbessi} B. Dandogbessi and A. Kenfack, Phys. Rev. E \textbf{92}, 062903 (2015).
\bibitem{Du} L. Du and D. Mei, Phys. Rev. E \textbf{85}, 011148 (2012).
\bibitem{Januszewski} M. Januszewski and J. Luczka, Phys. Rev. E \textbf{83}, 051117 (2011).
\bibitem{Ghosh} P. K. Ghosh, P. H\"{a}nggi, F. Marchesoni, and F. Nori, Phys. Rev. E \textbf{89}, 062115 (2014).
\bibitem{Malgaretti} P. Malgaretti, I. Pagonabarraga, and J. M. Rubi, Phys. Rev. Lett. \textbf{113}, 128301 (2014).

\bibitem{Marchesoni} F. Marchesoni, Phys. Lett. A \textbf{119}, 221 (1986).
\bibitem{Savel} S. Savel'ev, F. Marchesoni, P. H\"{a}nggi, and F. Nori, EPL \textbf{67}, 179 (2004).

\bibitem{Sarracino} A. Sarracino, F. Cecconi, A. Puglisi, and A. Vulpiani, Phys. Rev. Lett. \textbf{117}, 174501 (2016).
\bibitem{Cecconi} F. Cecconi, A. Puglisi, A. Sarracino, and A. Vulpiani, Eur. Phys. J. E \textbf{40}, 81 (2017).
\bibitem{Cecconi1} F. Cecconi, A. Puglisi, A. Sarracino, J. Phys.: Condens. Matter \textbf{30}, 264002 (2018).
\bibitem{Cividini} J. Cividini , D Mukamel, and H. A. Posch, J. Phys. A: Math. Theor. \textbf{51},  085001 (2018).

\bibitem{Haljas} A. Haljas, R. Mankin, A. Sauga, and E. Reiter, Phys. Rev. E \textbf{70}, 041107 (2004).

\bibitem{Slapik} A.Slapik, J.Luczka, and J.Spiechowicz, Commun Nonlinear Sci Numer Simulat \textbf{55}, 316 (2018).

\bibitem{Cleuren} B. Cleuren and C. Van den Broeck,  Phys. Rev. E \textbf{67} 055101(R) (2003).
\bibitem{Chen} R. Chen, W. Pan, J. Zhang, and L. Nie, Chaos \textbf{26}, 093113 (2016).

\bibitem{Li} J. H. Li and J. Luczka, Phys. Rev. E \textbf{82}, 041104 (2010).

\bibitem{Buceta} J. Buceta, J. M. Parrondo, C. V. Broeck, and F. J. Rubia, Phys. Rev. E \textbf{61}, 6287 (2000).

\bibitem{Mangioni} S. E. Mangioni, R. R. Deza, and H. S. Wio, Phys. Rev. E \textbf{63}, 041115 (2001).

\bibitem{Luo} J. Luo, K. A. Muratore, E. A. Arriaga, and A. Ros, Anal. Chem.  \textbf{88}, 5920 (2016).


\bibitem{Kuhlmann} H. C. Kuhlmann, M. Wanschura, and H. J. Rath, J. Fluid Mech. \textbf{336}, 267 (1997).
\bibitem{Solomon}T. H. Solomon and J. P. Gollub, Phys. Rev. A \textbf{38}, 6280 (1988).
\bibitem{Tabeling}P. Tabeling, Phys. Rep. \textbf{362}, 1 (2002).

\bibitem{Honeycutt} R. L. Honeycutt, Phys. Rev. A \textbf{45}, 600 (1992).

\bibitem{Costantini} G. Costantini and F. Marchesoni, EPL \textbf{48}, 491 (1999).
\bibitem{Reimann2} P. Reimann, C. Van den Broeck, H. Linke, P. H\"{a}nggi, J. M. Rubi, and A. Perez-Madrid, Phys. Rev. Lett. \textbf{87}, 010602 (2001).




\end{thebibliography}
\end{document}